\documentstyle[12pt]{article}
\begin{document}
\begin{flushright}

MRC-PH-TH-10-96

hep-th/9609177

\end{flushright}

\begin{center}

\title{Scattering from the Potential Barrier
$V=\cosh^{-2} \omega x$ from the Path Integration
over SO(1,2)} 
\author {H. Ahmedov$\ast$  and I.H.Duru$\ast\dag$ }
\maketitle
\end{center}
$\ast$ -TUBITAK -Marmara Research Centre, Research Institute for Basic
Siences, Department of Physics, P.O. Box 21, 41470 Gebze, Turkey

$\dag$ -Trakya University, Mathematics Department, P.O. Box 126, Edirne, Turkey.

\begin {center}
Abstract
\end{center}

Unitary irreducible representation of the group SO(1,2) is obtained in 
the mixed basis, i.e. between the compact and noncompact basis and the new 
addition theorems are derived which are required in path integral applications
involving the positively signed potential.
The Green function for the potential barrier $V=\cosh^{-2}\omega x$ is evaluated
from the path integration over  the coset space SO(1,2)/K where K is the
compact subgroup.The transition and the reflection coefficients are given.Results
for the moving barrier  $V=\cosh^{-2}\omega (x-g_0t)$ are also presented.

\

\

\

PACS number: 03.65.Db

\vfill
\eject

{\Large 1. Introduction}

Path integrations over the group manifolds or over the homogeneous spaces
are frequenly used for solving the path integrals of the quantum mechanical 
potentials \cite{kn:gros},\cite{kn:com},\cite{kn:per}.For example, P\"{o}schl-Teller, Hulthen and Wood-Saxon
potentials can be related to the group SU(2) \cite{kn:dur}.Path integrals over the
SU(1,1) manifold is studied for solving the modified P\"{o}schl-Teller
potential \cite{kn:bom}.

$V=-\cosh^{-2} \omega x $ i.e. the symmetric Rosen-Morse potential is
the special case of the potentials already mentioned in the above paragraph,
which can also be solved by transforming its Green function into the Green
function of the particle motion over the SO(3) manifold \cite{kn:ih}.

When the sign of the above potential is changed, that is when we consider
the well known potential barrier $V=\cosh^{-2} \omega x$, which is related
to the single soliton solution of KdV equation, special attention is required.
It is not the special case of the Rosen-Morse or the modified P\"{o}schl-
Teller potentials anymore. 
It is not related by coordinate transformation to them either.When one writes 
the e-value equation  for the Laplace-Beltrami (LB) operator in the 
space of matrix elements of the unitary irreducile representations of SO(1,2)
realized in the compact basis one arrives at the
Schr\"{o}dinger equation for the potential $V=\sinh^{-2} \omega x$ from 
which by the substitution $\omega x=\omega x'+i\frac{\pi}{2}$ we come to the 
potential $V=-\cosh^{-2} \omega x'$. To obtain the positive sign before
the latter potential one should diagonolize the LB operator
in the space of matrix 
elements of the unitary irreducible representation constructed in the
mixed basis i.e. between compact and non-compact basis. Such a necessity 
requries the derivation of new addition 
theorem for these matrix elements to get the path integral solution. 
In fact the construction of the unitary ireducible representations 
in the mixed basis
and the harmonic analysis on the double-sheeted hyperboloid in the hyperbolic 
coordinate system  are the basic
ingredients of the present note. The approach we adopt is of general nature 
which can be used to 
obtain the path integral solutions in any homogenous 
space in any parametrization.It leads to the path integral solution 
for the  new class of potentials (see sec. 2). 
The wave functions of these potentials correspond
to the matrix elements of the unitary irreducible representations in the basis
defined by the choice of the group decomposition.

In Section 2 we briefly rewiew the several possible decompositions of 
SO(1,2) relevant to the coordinates employed in the coset spaces which are 
double-sheeted and single- sheeted hyperboloids and the cone.

In Section 3 we  formulate the LB operator  for the coset space SO(1,2)/SO(2). 
Following the derivation of SO(1,2) matrix elements in the mixed basis, 
we diagonolize LB  operator, 
and then arrive at the Schr\"{o}dinger equation of 
the potential barrier $V=\cosh^{-2} \omega x$.
Normalized wave functions and spectrum are given.

In Section 4 we present the path integral formulation over the homogeneous
space SO(1,2)/SO(2). Starting from the short time interval Kernel  and
making use of the newly derived addition theorems we expand the short time 
interval Kernel in terms of the group matrix elements.

In Section V,we study the path integration for the potential barrier
 $V=\cosh^{-2} \omega x$.Transmittion and Reflection coefficients are given.
 Formulas for the barrier moving with a constant speed $g_0$, i.e. for
 $V=\cosh^{-2}\omega (x-g_0t)$, which is more relavent to the solitonic
 potential are also  presented.

\vfill
\eject

{\Large 2.Decompositions of the group SO(1,2) and the related quantum systems}

To express the group SO(1,2) in the decomposed forms the following 
 one parameter subgroups  can be employed:
\begin{eqnarray}
a=  \left( 
\begin{array}{ccc}
\cosh \alpha & 0 & \sinh \alpha \\
0 & 1 & 0 \\
\sinh\alpha & 0 & \cosh\alpha 
\end{array} 
\right) 
h=  \left( 
\begin{array}{ccc}
\cosh \beta & \sinh \beta & 0 \\
\sinh \beta & \cosh \beta & 0 \\
0 & 0 & 1  
\end{array} 
\right)  \nonumber \\
k=  \left( 
\begin{array}{ccc}
1 & 0 & 0 \\
0 & \cos \psi & -\sin \psi \\
0 & \sin \psi  & \cos \psi
\end{array} 
\right) 
n=  \left( 
\begin{array}{ccc}
1+\frac{t^2}{2} & t & \frac{t^2}{2} \\
              t & 1 & t \\
-\frac{t^2}{2} & -t & 1-\frac{t^2}{2}
\end{array} 
\right) 
\end{eqnarray}  
where
\begin {equation}
\alpha,\beta \in (-\infty,\infty), \ \ \ \ \ \psi \in (0,2\pi), \ \ \ \ \ 
t\in(-\infty ,\infty )
\end {equation}

G=SO(1,2) leaves the form  $(x,x)=x_0^2-x_1^2-x_2^2$  invariant. There are 
three possibilities:

\

(I) $\underline{Double - Sheeted \ hyperboloid \ M \ : \ (x,x) \ > \ 0}$
 \cite{kn:vil},\cite{kn:v}.

The vector $\dot{\xi}=(1,0,0)\in M$ is the stationary point of the 
compact subgroup k
as $\dot{\xi}k=\dot{\xi}$. Thus the decompositions of the group SO(1,2) related to 
the double-sheeted hyperboloid M is
\begin{equation}
g=kb,\ \  g\in SO(1,2),  \ \ k\in SO(2)
\end{equation}
The choice of the boost b defines the coordinate systems on M:

(i) b=ak or the Cartan decomposition of the group g=kak' defines 
the spherical 
coordinate parametrization of M. This choice is convenient for studying
the quantum mechanical potential $\frac{1}{\sinh^2\alpha}$.

(ii) b=ah or the non-compact Cartan decomposition of the group g=kah
defines the hyperbolic coordinates on M.It is suitable to the quantum mechanical 
system with potential $\frac{1}{\cosh^2\alpha}$. This case is the subject 
of the present  work.

(iii) b=an or the Iwasawa decomposition of the group g=kan
defines the parabolic coordinates on M which leads to the quantum mechanical 
system with potential $V=\exp(\alpha)$.

\

(II) $\underline{Single - Sheeted \ hyperboloid \ \overline{M} \ : \ (x,x) \ < \ 0}$
\cite{kn:vil},\cite{kn:v}

The vector $\dot{z}=(0,0,1)\in \overline{M}$ is the stationary point of the non-compact 
subgroup h as $\dot{z}h=\dot{z}$.The decompositions of the group SO(1,2) related to 
the single-sheeted hyperboloid $\overline{M}$ has the form 
\begin{equation}
g=hb,\ \  g\in SO(1,2),  \ \ h\in SO(1,1)
\end{equation}
with the possible choices of the boost b are given as the following:
  
(i) b=ak or the non-compact Cartan decomposition of the group g=hak defines 
the spherical coordinate parametrization of $\overline{M}$ which 
produces the potential  $-\frac{1}{\cosh^2\alpha}$.

(ii) b=$(aI^{\varepsilon} h,kI^{\varepsilon} h)$ or decomposition of 
the group 
g=$(h aI^\varepsilon h,h kI^\varepsilon h)$ \cite{kn:ver}
defines the hyperbolic coordinates on $\overline{M}$.
Here I is the metric tensor given by 
\begin {equation}
I=diag(1,-1,-1)
\end {equation}
and   $\varepsilon $=0,1. This decomposition is suitable to the quantum mechanical 
system with potentials $-\frac{1}{\sinh^2\alpha}$ and $-\frac{1}{\sin^2\phi}$.

(iii) $b=aI^\varepsilon n$ leads to the non-compact Iwasawa decomposition 
of the group $g=kaI^\varepsilon n $ and defines the parabolic coordinates on 
$\overline{M}$.The related quantum mecanical system is  $V=-\exp(\alpha)$.

\

 (III) $\underline{Cone \ M_0 \ : \ (x,x) \ = \ 0}$

The vector $\dot{y}=(1,0,1)\in M_0$ is the stationary point of the nilpotent 
subgroup n as $\dot{y}n=\dot{y}$.The decompositions of the group SO(1,2) related to 
the cone $M_0$ has the form $g=nb$ with b having the following forms :

(i) b=ak is the Iwasawa decomposition which defines spherical coordinates on $M_0$.

(ii) b=a$I^\varepsilon$h is the non-compact Iwasawa decomposition which
defines hyperbolic coordinates on $M_0$.

(iii) b=a$n^T$ is the Gauss decomposition which defines spherical coordinates on $M_0$

 It is impossible to relate quantum systems with the cone because the 
 metric tensor of $M_0$ is degenerate. This space is used for the 
 construction of the irreducible representations \cite{kn:gel}. 
 To construct the irreducible representations in the mixed basis
 we simultaneously have to use the realizations given by (i) and (ii) (see Appendix A).
\vfill
\eject

{\Large 3.The Double-Sheeted Hyperboloid in the
 Hyperbolic Coordinates}

We decompose the group G=SO(1,2) as 
\begin{equation}
g=hak
\end{equation}
Starting from the stationary point $\dot{\xi}=(1,0,0)$  we cover
all the homogeneous space M by the act  of the group elements as  
$x=\dot{\xi}g$.
Using (1) we get the parametrization of M
\begin {equation}
\xi=\dot{\xi}g=\dot{\xi}ah
=(\cosh \alpha \cosh \beta ,\cosh\alpha\sinh\beta ,\sinh\alpha )
\end {equation}

The metric tensor and the Laplace-Beltrami operator \cite{kn:hel} of M are
\begin {equation}
g_{M}=diag(-1,-\cosh^{2}\alpha),\ \ \ \ \   detg_{M}=\cosh^{2}\alpha
\end {equation}
and
\begin {equation}
\bigtriangleup=-\partial_{\alpha}^{2}-\tanh\alpha\ \partial_{\alpha}-
\cosh^{-2}\alpha\ \partial_{\beta}^{2}
\end {equation}

We will write down the e-value equation for the above LB operator in the space of SO(1,2) 
matrix elements which are evaluated between the compact and non-compact
basis.
The  matrix elements of the unitary principle series of the group SO(1,2)
in the mixed basis [see Appendix A]  given by
\begin {equation}
d_{\mu k}^{\sigma}(g)=\langle \mu \mid T^{\sigma}(g)\mid k \rangle 
\end {equation}
are the e-functions of the invariant differentiale operator $\bigtriangleup$
\begin {equation}
\bigtriangleup d_{\mu k}^{\sigma}(g)=-\sigma(1+\sigma)d_{\mu k}^{\sigma}(g)
\end{equation}
Here $\sigma$ is the weight of the representation
\begin {equation}
\sigma=-1/2+i\rho, \ \ \ \ \ \rho\in(0,\infty),
\end {equation}
and $\mid k \rangle\  and \mid \mu\rangle$ are the compact and non-compact basis corresponding to the
the degrees of freedom $\phi$ and $\beta$ respectfully.

Since we are dealing with the coset space M=G/K with K=SO(2), 
we do not need the full set of the matrix elements (10),
instead we employ
\begin{equation}
d_{\mu 0_k}^{\sigma}(ha)=\langle\mu\mid T^\sigma (ha)\mid 0_k\rangle
\end {equation}
Writing $d_{\mu 0_k}^\sigma (ha)$ as
\begin{equation}
d_{\mu 0}^\sigma (ha)=\exp (i\mu\beta )d_{\mu 0}^{\sigma}(a)
=\exp (i\mu\beta )         
(\det g_{M})^{-1/4} \Psi_{\mu}^{\sigma}(\alpha )
\end{equation}
the e-value equation (12) becomes
\begin {equation}
(-\partial_{\alpha}^{2}+(\mu^{2}+1/4)\cosh^{-2}\alpha+1/4)
\Psi_{\mu}^{\sigma}(\alpha)=
-\sigma(\sigma+1)
\Psi_{\mu}^{\sigma}(\alpha)
\end {equation}

which is equivalent to the Schr\"{o}dinger equation for the potential barrier
$V=\cosh^{-2}\alpha$ with an extra constant energy shift of $\frac{1}{4}$.
Note that if we would diagonolize the operator (9) in the space of 
matrix elements written between purely  compact basis,
the sign of potential would be 
negative.
The wave functions of the Sch\"{o}dinger equation (15) 
are given in terms of the Legendre functions 
\cite{kn:ar} by
\begin {equation}
\Psi_{\mu}^{\sigma}(\alpha)=
\frac{\cosh^{1/2}(\pi\rho)\cosh^{-1/2}(\alpha)}{\cosh(\pi\rho)
-isinh(\pi\mu)}
(P_{\sigma}^{i\mu}(i\sinh\alpha)+P_{\sigma}^{i\mu}(-i\sinh\alpha))
\end {equation}  
which are normalized as [see Appendix B]
\begin {equation}
\int_{-\infty}^{\infty}d\alpha  
\Psi_{\mu}^{\sigma} 
\overline{\Psi_{\mu'}^{\sigma'}}=
\frac{\delta(\mu-\mu')\delta(\sigma-\sigma')}{\rho\tanh\pi\rho}.
\end {equation}


\vfill
\eject

{\Large 4. Path Integration over the Coset Space M=G/K}

The probability amplitude 
for the particle of ``moment of inertia" m, to travel in the space M
from the point $\xi_{a}$ to $\xi_{b}$ 
in the time interval T is expressed by the path integral which in 
the time graded formulation is given by 
\begin{equation}
K(\xi_{a},\xi_{b};T)=\lim_{n\rightarrow \infty }
\int\prod_{j=1}^{n} d\xi_{j}
\prod_{j=1}^{n+1}K(\xi_{j-1},\xi_{j};\varepsilon)
\end{equation}
with $T=(n+1)\varepsilon$\ \ and \ \
$d\xi_{j}=(detg_{M})^{1/2}d\alpha_{j}d\beta_{j}$.

The Kernel connecting the points $\xi_{j-1}$ and $\xi_{j}$ 
separated by the small
time interval $t_{j}-t_{j-1}=\varepsilon$\ \  is \cite{kn:dur}
\begin{equation}
K_{j}=K(\xi_{j-1},\xi_{j};\varepsilon)=(\frac{m}
 {2i\pi\varepsilon})^{3/2}\exp(iS_{j})
\end{equation}
where $S_j$ is the short time interval action
\begin{equation}
S_j =\frac{m}{2\varepsilon} \delta^2_{j-1,j}
\end{equation}
The invariant distance between the points is 
\begin {equation}
\delta^{2}_{j,j-1}=
(\xi_{j}-\xi_{j-1},\xi_{j}-\xi_{j-1})=2-2\cosh\theta_{j-1,j}
\end {equation}
with
\begin{equation}
\cosh\theta_{j-1,j}=\cosh\alpha_{j-1}\cosh\alpha_{j}\cosh(\beta_{j-1}-\beta_{j})-
\sinh\alpha_{j-1}\sinh\alpha_{j}
\end{equation}

The short time interval Kernel (19) can be expanded in terms of the Legendre 
functions $P_{\sigma}(\cosh\theta)$ as ( with $\sigma=-1/2+i\rho$ )
\begin {equation}
K_{j-1,j}=
\int_{0}^{\infty}d\rho\rho\tanh(\pi\rho)C_{\sigma}
P_{\sigma}(\cosh\theta_{j-1,j})
\end {equation}
From the orthogonality condition
\begin {equation}
\int_{1}^{\infty}dzP_{\sigma}(z)\overline{P_{\sigma'}(z)}=\frac{1}
{\rho\tanh\pi\rho}\delta(\sigma-\sigma')
\end {equation}
the coefficients $C_\sigma$ is obtained :
\begin{equation} 
C_{\sigma}=-\frac{m}{\surd\pi\varepsilon}\exp(-\frac{i\pi}{\varepsilon})
K_{i\rho}(-\frac{m}{\varepsilon})
\end{equation}
Here $K_{i\rho}$ is the MacDonald function. In $\varepsilon\rightarrow 0$
limit,\ \ by using the asymptotic form of the MacDonald function \cite{kn:jun}
 we can write the short time interval Kernel (19) as
\begin{equation}
K_{j}\simeq\int_{0}^{\infty}d\rho\rho\tanh\pi\rho
\exp(\frac{i\varepsilon}{2m}\sigma(\sigma+1)) P_{\sigma}(\cosh\theta_{j})
\end{equation}
By the help of the addition theorem [see Appendix C] for the complete set of functions on the
homogeneous space M we get
\begin{equation}
K_j=\int_{-\infty}^\infty d\mu\int_0^\infty d\rho\rho\tanh\pi\rho
\exp(\frac{i\varepsilon}{2m}\sigma(\sigma+1))
d_{\mu 0}^\sigma(\xi_{j-1})
\overline{d_{\mu 0}^\sigma (\xi_j)}
\end{equation}
We first insert the above form of the short time interval Kernel into (18);
then by making use of 
the orthogonality condition (17) we can execute the $\prod_{j-1}^n d\xi$
integrals:
\begin{eqnarray}
K(\xi_a ,\xi_b;T)=\int_{-\infty}^\infty d\mu\int_0^\infty
d\rho\rho\tanh\pi\rho \nonumber \\
\exp(-\frac{i(\rho^2+1/4)}{2m}T)
d_{\mu 0}^\sigma(\xi_a)\overline{d_{\mu 0}^\sigma(\xi_b)}
\end{eqnarray}
By using the addition theorem the  Kernel (28) can be written in the form
\begin{equation}
K(\xi_a ,\xi_b;T)=\int_0^\infty 
d\rho\rho\tanh\pi\rho\exp(-i\frac{\rho^2+1/4}{2m}T)
P_\sigma(\cosh\theta_{ab})
\end{equation}
where $\cosh\theta_{ab}$ depends on the coordinates of the points a and b 
through the relation defined by (22).

The Fourier transform of (29) can be calculated to obtain the energy
dependent Green function $G(\xi_a ,\xi_b ;E)$
\begin{equation}
G(\xi_a ,\xi_b ;E)=\int_0^\infty\exp(iET) K(\xi_a ,\xi_b;T)
=2mQ_{-1/2-i\surd(2mE-\frac{1}{4})}(\cosh\theta_{ab})
\end{equation}
where Q is the Legendre function of the second kind.
In deriving (30) we used the connection between the Legendre functions
of the first and second kind \cite{kn:grad}
\begin{equation}
\int_0^\infty dx\frac{x\tanh\pi x}{a^2+x^2}P_{-1/2+ix}(z)=Q_{a-1/2}(z)
\end{equation}

\vfill
\eject

{\Large 5.Path Integral for the Potential Barrier  $V=V_0\cosh^{-2}(\omega x)$}

Phase space path integral for the particle of mass m moving under the influence 
of the potential barrier $V=V_0\cosh^{-2}(\omega x)$ is
\begin{equation}
K(x_a ,x_b;T)=\int Dx Dp_x 
\exp(i\int_0^T dt(px-p_x^2/2m-V_0\cosh^{-2}\omega x))
\end{equation}
which is in the time graded formulation equal to
\begin{eqnarray}
K(x_a ,x_b;T)=\lim_{n\rightarrow\infty}\int\prod_{j=1}^ndx_j
\prod_{j=1}^{n+1}d\frac{p_{xj}}{2\pi}
\prod_{j=1}^{n+1} \nonumber \\
\exp[i(p_{xj}(x_j-x_{j-1})-\varepsilon p_{xj}^2/2m-
\varepsilon V_0\cosh^{-2}\omega x_j)]
\end{eqnarray}
The phase space formulation (32) easily enable us to establish the connection
between our quantum mechanical problem and the path integration over
the coset space M=SO(1,2)/K.
In fact when we consider the Hamiltonian $H_M$ for the particle motion over
 the coset space M=G/K (recall the Sch\"{o}dinger equation (15)):
\begin{equation}
H_M -\frac{1}{4} \Rightarrow\frac{1}{2m}(p_\alpha^2+p_\beta^2\cosh^{-2}\alpha)
-\frac{1}{4}(\frac{\omega^2}{2m})
\end{equation}
we observe that the Hamiltonian in the action of (32) resembles the above Hamiltonian with
the momentum $p_\beta$ fixed to the value $ p_\beta=\surd(2mV_0) $.
In writing (34) we introduced the corrections due to parameters $\omega$
and (2m) which were equal to 1 in sections 3 and 4.
 Thus we can convert 
the path integral (32) into the path integral for the motion in the space
M=G/K.
We first rescale x by $ \omega x=\alpha $ ;\ \ \ $ p_x=\omega  p_\alpha $ and
arrive at:
\begin{equation}
K(x_a ,x_b;T)=\omega\int D\alpha Dp_\alpha\exp[i\int_0^{\omega^2 T}dt
(p_\alpha\alpha
-p_\alpha^2/2m-V_0\cosh^{-2}\alpha)]
\end{equation}
We then rewrite the potential term in the above path integral by introducing 
an auxilary dynamics by extending the phase space with the identity
\begin{eqnarray}
\exp[-i(\int_0^{\omega^2 T} dtV_0\cosh^{-2}\alpha)]=
\int d\beta_b\exp(-\surd(2mV_0/\omega^2)(\beta_b-\beta_a)) \nonumber \\
\lim_{n\rightarrow\infty}\int\prod_{j=1}^nd\beta_j
\prod_{j=1}^{n+1} d\frac{p_{\beta j}}{2\pi}
\prod_{j=1}^{n+1}
\exp(i(p_{\beta j}(\beta_j-\beta_{j-1})-
\frac{\omega^2\varepsilon p_{\beta_j}^2}{2m\cosh^2\alpha})
\end{eqnarray}
which can be proven by direct calculation. Note that the phase space formulation is rather 
essential for the above identity which establishes the connection between 
our quantum mechanical problem and the partical mation over M.
The identity of (36) enables us to reexpress (35) as
\begin{equation}
K(x_a ,x_b;T)=\omega\int d\beta_b\exp(-\surd(2mV_0/\omega^2)(\beta_b-\beta_a))
\exp(i\frac{\omega^2}{8m}T) K_M(\xi_a ,\xi_b;T)
\end{equation}
where $ K_M $ is the Kernel for the motion over the manifold M=G/K which is 
studied in the previous chapter. The factor $\exp(i\frac{\omega^2}{8m}T)$
in the above equation reflects (see eqs. (15) and (34)) the 1/4 energy
 difference between the potential barrier $\cosh^{-2}omega x$ and the
 particle motion over the coset space M.
 We then insert the expression (28) into (37),
use (14) for the matrix elements $ d_{\mu0}^\sigma $ and arrive at
\begin{eqnarray}
K(x_a ,x_b;T)=2\pi\omega(\cosh\omega x_a\cosh\omega x_b)^{-1/2} \nonumber \\
\int d\rho\rho\tanh\pi\rho exp(-i\frac{\rho^2\omega^2}{2m}T)
\psi^\sigma_{\frac{(2mV_0)^{1/2}}{\omega}}(\omega x_a)
\overline{\psi^\sigma_{\frac{(2mV_0)^{1/2}}{\omega}}(\omega x_b)}  
\end{eqnarray}
which displays the wave functions.The asymptotic form of the wave functions 
are
\begin{equation}
\lim_{x\rightarrow\infty} 
\psi^\sigma_{\frac{(2mV_0)^{1/2}}{\omega}}(\omega x)
\simeq
\frac{\Gamma (1/2-i\rho)}
{\Gamma (1/2-i(\rho + \frac{(2mV_0)^{1/2} }{\omega}))}
\exp(-i\rho\omega x)
\end{equation}
\begin{equation}
\lim_{x\rightarrow  -\infty}
\psi^\sigma_{\frac{(2mV_0)^{1/2}}{\omega}}(\omega x)
\simeq 
\frac{\Gamma (1/2-i\rho)}
{\Gamma (1/2-i(\rho + \frac{(2mV_0)^{1/2} } {\omega}))}
(T\exp(-i\rho\omega x)+R\exp(i\rho\omega x))
\end{equation}
in which the transition and the reflection coefficients are identified as:
\begin{equation}
T=\frac
{\Gamma (1/2+i(\rho + \frac{(2mV_0)^{1/2}}{\omega} ))
\Gamma (1/2+i(\rho - \frac{(2mV_0)^{1/2}}{\omega}))}
{\Gamma (i\rho)\Gamma(1+i\rho)}
\end{equation}
\begin{equation}
R=\frac
{\Gamma (1/2+i(\rho + \frac{(2mV_0)^{1/2}}{\omega} ) )
\Gamma (1/2+i(\rho - \frac{(2mV_0)^{1/2}}{\omega}) )
\Gamma (-i\rho)}
{\Gamma (1/2+i \frac{(2mV_0)^{1/2}}{\omega} )
\Gamma (1/2-i\frac{(2mV_0)^{1/2}}{\omega} )
\Gamma (i\rho)}
\end{equation}
If one considers the same potential barrier in motion with a constant
speed
\begin {equation}
V(x,t)=\frac{V_0}{\cosh^2 \omega (x-g_0t)} \ \ ; \ \ \ \ g_0=const.
\end{equation}
the Kernel becomes  \cite{kn:hak}:
\begin{equation}
K_{g_0}(x_a,x_b;T)=\exp(-i\frac{m}{2}g_0^2T)\exp(-img_0(x_b-x_a))
K(x_a-g_0t_a,x_b-g_0t_b:T)
\end{equation}
Here the form of K is given by (38).From the above formula the wave functions
are recognized as
\begin{equation}
\psi_{g_0}(x,t)=\exp(-i\frac{m}{2}g_0^2t)\exp(-img_0x)
\psi^\sigma_{\frac{(2mV_0)^{1/2}}{\omega}}(\omega (x-g_0t))
\end{equation}
where $\psi(\omega (x-g_0t))$ is obtained from the static one
given in (16) and (38) by simply replacing $x$ by $x-g_0t$.
For finite values of time variable $t$ the transition and the reflection
coefficients remain in the static forms of (41) and (42).

Inspecting the limiting forms of T and R we obtains
\begin{equation}
 \mid\frac{T}{R}\mid\rightarrow\rightarrow\infty  \ as \ \rho\rightarrow\infty, \ and \ 
 \mid\frac{T}{R}\mid\rightarrow0 \ as \ \rho\rightarrow 0
\end{equation}
We see that the low energy waves are mostly reflected, while the high energy waves
are more easily transmitted through the barrier.

Inspecting (47) and the asymptotic forms as $x\rightarrow\infty$ we observe 
that the barrier motion contributes the following constant additional term 
to the energy
\begin{equation}
\bigtriangleup E=\frac{mg_{0}^2}{2}-\rho\omega g_0
\end{equation}
The first term of the above extra energy is of the kinetic energy type 
(for $g_0$ has the dimension of velocity). It is also interesting that the 
barrier motion introduces the extra ondulation of Doppler nature trough
$\exp(-img_0x)$ term in the wave function.

\vfill
\eject

\begin {center}
\huge {Appendix}
\end {center}

{\Large A. The Unitary Irreducible Representations of the Group SO(1,2)
in the Mixed Basis}
\renewcommand{\theequation}{A.\arabic{equation}}
\setcounter{equation}{0}

We will construct the irreducible representations of the pseudo-orthogonal
group G=SO(1,2) in the space of the infinitely differentiable 
homogeneous functions $F(y)$ with the homogenity degree $\sigma$ on 
the cone $Y:\ \  [y,y]=0 $. 
\begin {equation}
T^\sigma (g)F(y)=F(yg),\ \ \ \ \ \     g\in G ,\   y\in Y ,
\end {equation}
\begin {equation}
F(ay)=a^\sigma F(y) ,\ \ \ \ \ \    a\in R ,\   \sigma\in C
\end {equation}

In order to construct the matrix elements of the representation in the mixed 
basis we have to define the cone in two coordinate systems corresponding
to these basis. Use the Iwasawa decompositions for the group SO(1,2) 
\cite{kn:vil}:
\begin{equation}
g=n(t)a(\theta )k(\phi ) ,\ \ \ \ \   g\in G
\end{equation}
\begin{equation}
g=n(t)I^\varepsilon a(\gamma )h(\beta ) ,\ \ \ \ \   g\in G
\end{equation}

Here the element n of the nilpotent subgroup and other matrices $a,h,k$
are the ones defined in (2).
The stationary point of the nilpotent subgroup is $\dot{y}=(1,0,1)$ 
\begin {equation}
\dot{y}n(t)=\dot{y},
\end {equation}

The coset space Y=G/N is equivalent to the cone Y given by $y=\dot{y}g$
which can be defined in two realizations as:
\begin{equation}
y=\exp(\gamma_k )s_k ,\ \ \  s_k=\dot{y}k(\phi )=(1,\sin\phi ,\cos\phi)
\end{equation}
\begin{equation}
y=\exp(\gamma_h )s_h ,\ \ \  s_h=\dot{y}h(\beta )=
(\cosh\beta ,\sinh\beta ,(-1)^\varepsilon )
\end{equation}
The connection between the above realizations is:
\begin{equation}
\exp(\gamma_h )s_h=\exp(\gamma_k )s_k
\end{equation}
or
\begin{equation}
\cosh\beta\exp(\gamma_h )=\exp(\gamma_k ),\ \ \ \cos\phi =\frac{(-1)^\varepsilon}
{\cosh\beta},\ \ \ \sin\phi=\tanh\beta
\end{equation}
Using (A.2) we get:
\begin{equation}
F(y)=\exp(\gamma_h\sigma)F(s_h)=\exp(\gamma_k\sigma)F(s_k)
\end{equation}
We know that  the principal series of the irreducible representation in the
compact basis (the group decomposition is $g=kak$ ) is
unitary with respect to the scalar product
\begin{equation}
(F_1,F_2)=\frac{1}{2\pi}\int_{0}^{2\pi} d\phi F(s_k)\overline{F(s_k)} 
\end{equation}
if  $\sigma=-1/2+i\rho ,\rho\in(0, \infty) $ [11].
Using the relation (A.10) we can introduce the scalar product in the
space of representation in the  mixed basis:
\begin{equation}
\langle F_1,F_2\rangle=\frac{1}{2\pi}\int_0^{2\pi} d\phi\overline{\exp(
(\gamma_h-\gamma_k)\sigma ))}\overline{F(s_h)}F(s_k)
\end{equation}
The invariant bilinear Hermitian form (A.12) can be written as
\begin{equation}
\langle F_1,F_2\rangle=\frac{1}{2\pi}\int_{-\infty}^{\infty} d\beta
\cosh\beta^\sigma \overline{F(s_h)}F(s_k)
\end{equation}
From (A.1) we get the representation formulas corresponding to the above
mentioned realizations:
\begin {equation}
T^\sigma (g)F(s_k)=\exp((\gamma^g_k-\gamma_k)\sigma )F(s_{k^g}) ,\ \ \ \ \  g\in G
\end {equation}
\begin {equation}
T^\sigma (g)F(s_h)=\exp((\gamma^g_h-\gamma_h)\sigma )F(s_{h^g}) ,\ \ \ \ \  g\in G
\end {equation}
Here $s_{k^g}$ \ \ and $s_{h^g}$ are defined  as
\begin {eqnarray}
\exp(\gamma^g_k ) s_{k^g}=\exp(\gamma_k ) s_kg \nonumber \\
\exp(\gamma^g_h ) s_{h^g}=\exp(\gamma_h ) s_hg
\end {eqnarray}
We see that the natural representations for the maximal compact k=SO(2) and
noncompact h=SO(1,1) subgroups corresponding to the realizations of 
the representations (A.14) and (A.15) are
\begin {equation}
T^\sigma (k(\phi_0))F(s_{k(\phi)})=F(s_{k(\phi +\phi_0)})
\end {equation}
and
\begin {equation}
T^\sigma (h(\beta_0)F(s_{h(\beta)})=F(s_{h(\beta +\beta_0)})
\end {equation}
By the help of the  expansion formulas
\begin {eqnarray}
F(s_k)=\sum_{n=-\infty}^\infty C_n\exp(in\phi)  \nonumber \\
F(s_h)=\int_{-\infty}^\infty d\mu C_\mu\exp(i\mu\beta)  
\end {eqnarray}
we can rewrite (A.17)and (A.18) as
\begin {eqnarray}
T(k(\phi_0))\exp(in\phi)=\exp(in\phi_0)\exp(in\phi) \nonumber \\
T(h(\beta_0)\exp(i\mu\beta)=\exp(i\mu\beta_0)\exp(i\mu\beta)
\end {eqnarray}
which coincide with the unitary irreducible representations of the subgroups
SO(2) and SO(1,1).

Now we are ready to construct the unitary irreducible representation
for the group SO(1,2) in the mixed basis. Let us introduce the function D(g)
\begin{equation}
D(g)=\langle F_1\mid T^\sigma (g)\mid F_2\rangle
\end{equation}
in terms of the Hermitian bilinear form  given by (A.13).
Using the expansion formulas (A.19) we obtain
\begin{equation}
D(g)=\sum_{n=-\infty}^{\infty}\int_{-\infty}^\infty d\mu C_n \overline
{C_\mu} d_{\mu n}^\sigma (g)
\end{equation}
where $d_{\mu n}^\sigma (g)$ are the matrix elements of the unitary irreducible
representation
\begin{equation}
d_{\mu n}^\sigma (g)=\langle\mu\mid T^\sigma (g)\mid n\rangle,\ \ \ \ g=hak\in G
\end{equation}
By the help of the group property $T^\sigma (hak)=T(h)T^\sigma (a)T(k)$ and
the expressions in  (A.20) we obtain
\begin{equation}
d_{\mu n}^\sigma (g)=\exp(-i\mu\beta )\langle\mu\mid T^\sigma (a)
\mid n\rangle\exp(in\phi)
\end{equation}
The integral representation for the matrix elements  corresponding to
the subgroup $a(\alpha )$ is (in the case n=0)
\begin {equation}
d_{\mu 0_k}^{\sigma}(a)=\sum_{\varepsilon=0}^{1}
\int_{-\infty}^\infty d\beta   
\exp(-i\mu\beta ) (\cosh\beta\cosh\alpha+(-1)^\varepsilon\sinh\alpha )^
\sigma
\end {equation}
Evaluting this integral we get:
\begin {equation}
d_{\mu 0_k}^{\sigma}(a(\alpha))=\frac{\cosh^{1/2}(\pi\rho)}
{\cosh(\pi\rho)-isinh(\pi\mu)}
(P_{\sigma}^{i\mu}(i\sinh\alpha)+P_{\sigma}^{i\mu}(-i\sinh\alpha))
\end {equation}

{\Large B. The Orthogonality Condition}
\renewcommand{\theequation}{B.\arabic{equation}}
\setcounter{equation}{0}

Consider the expression
\begin{equation}
B^{\sigma\sigma '}_{\mu\mu '}=\int_G dg d^\sigma_{\mu
0_k}(g)\overline {d^{\sigma '}_{\mu '0_k}(g)} \end{equation}
with g=hak. We first  change the variables in the above
integral
\begin{equation} hak=k'a'k''
\end{equation}
which is equivalent to passing from $g=hak$ to the Cartan
decomposition $g'=k'a'k''$ \cite{kn:vil}. Using the completness
condition for the matrix elements of the maximal compact
subgroup K=SO(2):
\begin{equation}
\sum_{n=-\infty}^{\infty}\mid n\rangle \langle n\mid=1
\end{equation}
 and the equality
\begin{eqnarray}
d^\sigma_{\mu 0_k}(g)=\langle\mu\mid T^\sigma (g)\mid
0_k\rangle= \sum_{n=-\infty}^\infty\langle\mu\mid n\rangle
\langle n\mid T^\sigma (g)\mid 0_k\rangle= \nonumber \\
=\sum_{n=-\infty}^\infty\langle\mu\mid n\rangle d^\sigma_{n0_k}(g)
\end{eqnarray}
we get
\begin{eqnarray}
B^{\sigma\sigma '}_{\mu\mu '}=
\sum_{n,n'=-\infty}^{\infty} \langle \mu \mid n \rangle
\langle \mu ' \mid n ' \rangle
\int_{G'} dg' d^\sigma_{n0_k}(g')\overline{d^{\sigma '}_{n' 0_k} (g')}  \nonumber \\
=\sum_{n,n'=-\infty}^{\infty}\langle\mu\mid n\rangle
\overline{\langle\mu '\mid n'\rangle} \frac{\delta(\rho-\rho
')\delta_{n n'}}{\rho\tanh\pi\rho}= \frac{\delta(\rho-\rho
')\delta (\mu -\mu ')}{\rho\tanh\pi\rho}
\end{eqnarray}
In (B.5) we used the orthogonality condition of the matrix
elements in the Cartan basis \cite{kn:vil}, \cite{kn:ver}. together with the
orthogonality condition for the matrix elements of the
maximal noncompact subgroup  SO(1,1)
\begin{equation}
\langle\mu\mid\mu '\rangle=\delta(\mu-\mu ')
\end{equation}
It is obviously equivalent to
\begin{equation} \int_G dg d^\sigma_{\mu
0_k}(g)\overline{d^{\sigma '}_{\mu '0_k}(g)}=
\frac{\delta(\rho-\rho ')\delta (\mu -\mu
')}{\rho\tanh\pi\rho} \end{equation} where the invariant
measure is $dg=(det_m g)^{1/2}d\alpha d\beta d\phi$

Taking into account (14) we get the orthogonality condition for the wave
functions:
\begin {equation}
\int_{-\infty}^{\infty}d\alpha  
\Psi_{\mu}^{\sigma} 
\overline{\Psi_{\mu'}^{\sigma'}}=
\frac{\delta(\mu-\mu')\delta(\sigma-\sigma')}{\rho\tanh\pi\rho}
\end {equation}

{\Large C. The Addition Teorem and the Completeness Condition}
\renewcommand{\theequation}{C.\arabic{equation}}
\setcounter{equation}{0}

Note that the Legendre functions appear as the zonal spherical functions
of the representation of the group SO(1,2) if the elements of the group G
has the Cartan decomposition $g=kak'$:
\begin {equation}
P^\sigma (\cosh\theta )=d^\sigma_{0_k,0_k}(a(\theta ))
\end {equation}
Suppose that  $g_1$ and $g_2$ are the elements of the group G which have the following
decompositions 
\begin {equation}
g_j=h_ja_jk_j,\ \  j=1,2 \ \ and\ \  g_{12}=h_{12}a_{12}k_{12}=g_1^{-1}g_2 
\end{equation}
Write the element $g_{12}$ in the Cartan decomposition: 
\begin{equation}
g_{12}= k'_{12}a'_{12}k_{12} 
\end{equation}
Using (C.1),(C.2) and (C.3) we obtain:
\begin {eqnarray}
P^\sigma (\cosh\theta_{12})=d^\sigma_{0_k,0_k}(a(\theta_{12}))=
d^\sigma_{0_k,0_k}(k'_{12}a(\theta_{12})k_{12})= \nonumber \\
=d^\sigma_{0_k,0_k}(g^{-1}_1g_2)=\int_{-\infty}^\infty d\mu d^\sigma_{0_k,\mu}(g^{-1}_1)
d^\sigma_{\mu ,0_k}(g_2)
\end {eqnarray}
Making use of the property of the matrix elements 
\begin {equation}
d^{-\sigma-1}_{0_k,\mu}(g^{-1})= \overline{d^\sigma_{\mu ,0_k}(g)}
\end {equation}
and the equivalence of the representations $T^\sigma$ and $T^{-\sigma-1}$ 
\cite{kn:ver} we get the final result:
\begin {equation}
P^\sigma (\cosh\theta_{12})=
\int_{-\infty}^\infty d\mu \overline{d^\sigma_{\mu,0_k}(g_1)}
d^\sigma_{\mu ,0_k}(g_2)
\end {equation}
Here $\cosh\theta_{12}$ is defined from the algebraic equation (C.3) and
is given by 
\begin{equation}
\cosh\theta_{1,2}=\cosh\alpha_{1}\cosh\alpha_{2}
\cosh(\beta_{1}-\beta_{2})-
\sinh\alpha_{1}\sinh\alpha_{2}
\end{equation}
which coincide with (22).

The completness condition for the matrix elements
on the homogeneous space $g\in M$ are given by
\begin{equation}
\int_0^\infty d\rho\rho\tanh\pi\rho\int_{-\infty}^\infty d\mu
d^\sigma_{\mu 0_k}(g)\overline{d^\sigma_{\mu 0_k}(g')}=\delta (g-g')
\end{equation}
To prove the above relation one considers the connection
between the invariant differential operator ( Laplace-Beltrami operator ) on the
manifold M with the Schr\"{o}dinger equation.Since the Schr\"{o}dinger equation
has only the continuous specrum, the spectrum of the
invariant differential operator should also be continuous. 
Therefore  the discrete unitary
series of the irreducible representation does not make contribution and can
be ignored. From the physical point of view  the complementary series can also     
be ignored.

\vfill
\eject

\end{document}